# Universal effect of ammonia pressure on synthesis of colloidal metal nitrides in molten salts

Ruiming Lin,[1] Vikash Khokhar,[2] Ningxin Jiang,[3] Wooje Cho,[3] Zirui Zhou,[3] Di Wang,[3] Justin C. Ondry,[3] Zehan Mi,[1] James Cassidy,[3] Alex M. Hinkle,[3] Alexander S. Filatov,[3] John S. Anderson,[3] Richard D. Schaller,[4,5] De-en Jiang,[6] Dmitri V. Talapin[1,3,5,*]

[1] *Pritzker School of Molecular Engineering, University of Chicago, Chicago, IL 60637, USA*

[2] *Interdisciplinary Materials Science Program, Vanderbilt University, Nashville, TN 37235, USA*

[3] *Department of Chemistry and James Franck Institute, University of Chicago, Chicago, IL 60637, USA*

[4] *Department of Chemistry, Northwestern University, Evanston, IL 60208, USA*

[5] *Center for Nanoscale Materials, Argonne National Laboratory, Argonne, IL 60517, USA*

[6] *Department of Chemical and Biomolecular Engineering, Vanderbilt University, Nashville, TN 37235, USA*

*Correspondence to: dvtalapin@uchicago.edu

**Metal nitrides represent a large class of materials with extensive applications in optoelectronics, energy, and healthcare technologies. For example, GaN and related nitride semiconductors are key materials for solid-state lighting and high-power electronics[1,2]; TiN and other early transition metal nitrides (TMNs) are widely used in wear-resistant alloys, tool coatings, catalysts and medical implants[3]. Strong metal–nitrogen bonds grant nitrides structural rigidity as well as chemical and thermal stability[4]. However, the covalency of metal-nitrogen bonds necessitates high temperatures to synthesize crystalline metal nitrides. Common synthetic routes include high-temperature solid-state nitridation[5], crystal growth in supercritical ammonia[6], molecular-beam epitaxy (MBE)[7], reactive sputtering[8,9], and chemical vapor deposition (CVD)[1,10,11,12]. The solution synthesis of colloidal nitride nanocrystals (NCs) is rare and particularly challenging because commonly used solvents and surfactants decompose at temperatures far below those required for crystallization of most metal nitrides. Here we report a general approach to solution synthesis of colloidal metal nitride NCs by reacting metal halides and ammonia dissolved in molten inorganic salts at**

**elevated pressures. Successful syntheses of colloidal TiN, VN, GaN, NbN, $Mo_2N$, $Ta_3N_5$, $W_2N$, as well as ternary $Ti_{1-x}V_xN$ NCs, are demonstrated. These NCs expand the scope of solution-processable technologically important materials.**

In recent years, significant progress has been made with synthesizing functional materials in the form of colloidal NCs that combine solution processability, size-dependent optical properties of semiconductor quantum dots (QDs) and plasmonic materials, and high surface-to-volume ratios important for catalytic applications[13,14]. Previously, some nitride nanoparticles have been synthesized by solid-state nitridation of oxides, laser ablation or plasma methods[15,16,17,18,19], but these approaches offer only a limited ability to control phase purity, particle size and dispersibility in solvents. In contrast, solution-synthesized colloidal nitride NCs have the potential to combine the unique material properties of nitrides with the scalability and tunability of colloidal systems. Despite this promise, solution syntheses of nitride NCs have seen only limited success especially when compared to well-established colloidal materials such as CdSe and InP[20,21]. Some successful examples include thermodynamically metastable and/or barely stable late transition metal nitrides such as $Ni_3N$, $Cu_3N$, $Cu_3PdN$ and $Zn_3N_2$[22,23,24,25,26]. The synthetic difficulty of nitride NCs can be illustratively explained by comparing the standard enthalpies of formation for TiN (-265.8 kJ/mol) and InP (-88.7 kJ/mol) phases – this stark thermodynamic difference is a direct result of the high strength and covalent nature of metal-nitrogen bonds[27]. The growth of a defect-free crystalline lattice generally requires microreversibility of elemental steps to eliminate occasional structural defects. Synthesizing refractory materials with strong bonds typically requires high temperatures to access this microreversibility. However, for colloidal synthesis, maximal temperatures are limited by the thermal stability of the solvents and surfactants used, which is below ~400 °C for organic compounds[21]. Alternatively, molten inorganic salts have been recently introduced as a medium to provide high-temperature colloidal stability of NCs[28] and thus enable the synthesis of previously inaccessible Group III phosphide, arsenide, and antimonide NCs[29,30,31]. Along parallel lines, it has been shown that molten salts can also facilitate the formation of nanocrystalline powders of boride, carbide and nitride phases[32,33,34,35,36]. This motivates our investigations into the use of molten salts to similarly generate colloidal nitride NCs.

Metal nitrides are often synthesized by utilizing ammonia as a convenient and inexpensive nitrogen source. The phase diagram of ammonia is shown in **Fig. 1a**, superimposed on typical conditions used in different nitride synthesis methods. Thus, CVD synthesis of GaN and other nitrides typically requires temperatures exceeding 500 °C[1,10,11,12], while the ammonothermal growth of bulk GaN crystals is carried out around 600 °C and 200 MPa in supercritical ammonia[6]. We therefore hypothesized that using elevated pressures of ammonia in molten salt syntheses might be a viable path to nitride NCs. In this study, we find that many metal nitrides can be synthesized in form of colloidal NCs at 425-600 °C by applying ammonia pressure of 1-5 MPa (**Fig. 1a**). This high-temperature elevated-pressure synthetic space is inaccessible for traditional solution synthesis in organic solvents and has not been previously explored by colloidal chemists.

In a typical synthesis (**Fig. 1b**), we used Lewis-basic alkali metal halides as solvents that simultaneously provide high-temperature stability, high solubility of metal halide precursors ($TiI_4$, $VCl_3$, $Ga_2I_4$, $NbCl_5$, *etc.*), and support the colloidal stability of nucleated nitride NCs. Metal halides were mixed with a finely ground CsI/KI/NaI eutectic powder (molar ratio 52:4:44, melting point 407 °C) to form a ~1 *mol L*$^{-1}$ solution. The mixture was then added to a high-pressure reactor (Supplementary Fig. 1) and heated to a desired temperature (450-600 °C) under a dry nitrogen atmosphere to form a homogeneous solution (Supplementary Figs. 2 and 3). High-pressure $NH_3$ was then injected into the reactor, and the temperature was maintained for 5 min. The reaction was then cooled down to room temperature, and the products were retrieved from the salt matrix with methanol which dissolved the alkali and ammonium halides. Stable colloidal solutions were obtained by subsequent dispersion of the nitride NCs in non-polar organic solvents such as toluene with the addition of oleic acid and oleylamine as surface ligands (**Fig. 1c**).

The ammonia pressure, together with the reaction temperature, are key parameters for controlling the reaction products. For example, the reaction of $VCl_3$ with $NH_3$ at 500 °C yields three distinctive outcomes depending on the ammonia partial pressure. At 100 kPa, nanocrystalline cubic-phase VN powder consisting of sintered grains with an average Scherrer size of 6.1 nm forms (**Fig. 1d** and Supplementary Fig. 4). When the $NH_3$ pressure is increased to 2.0 MPa, we observe the formation of discrete and monodisperse VN NCs. The characterizations of these NCs with Transmission Electron Microscopy (TEM, **Figs. 1e** and **2a**), powder X-ray diffraction (XRD) and small-angle X-ray scattering (SAXS, Supplementary Fig. 5) revealed cubic phase (space group 225) VN NCs with well-defined cuboid shapes, Scherrer size of 6.8 nm and a ~8.8% size distribution. These VN NCs form stable colloidal solutions in toluene (Supplementary Fig. 6).

Importantly, this synthesis uses simple reactants and achieves a ~75% reaction yield of colloidal NCs (Supplementary Table 1). Further increasing the NH$_3$ pressure to 5.0 MPa results in an increase of the Scherrer size of the VN crystalline domains to 16.8 nm (**Fig. 1f**) and this trend continues with increasing NH$_3$ pressure, ultimately preventing the formation of stable colloidal dispersions. A very similar NH$_3$ pressure dependence was observed for GaN (Supplementary Fig. 7).

This synthetic approach to colloidal metal nitrides is surprisingly universal – multiple binary (TiN, VN, GaN, NbN, Mo$_2$N, Ta$_3$N$_5$, W$_2$N) and ternary (*e.g.*, Ti$_{1-x}$V$_x$N, Supplementary Fig. 8) nitrides successfully nucleate and grow as discrete colloidal NCs within the pressure and temperature region highlighted with blue in **Fig. 1a**. We summarize the reaction conditions, powder XRD patterns, and TEM images of nitride NCs synthesized in molten salts in **Fig. 2**. Colloids of cubic rock salt phase TiN NCs were produced using temperatures above 500 °C with 0.8 MPa of NH$_3$ pressure (**Fig. 2b**). Colloidal GaN NCs with a hexagonal wurtzite phase were produced using 2.0 MPa of NH$_3$ at 450 °C (**Fig. 2c**); however, both wurtzite and zinc blende polytypes could be synthesized by small variations of the reaction conditions (Supplementary Figs. 7 and 9). Higher temperatures are necessary to grow NbN NCs (**Fig. 2d**) and Mo$_2$N NCs (**Fig. 2e**). Slight shifts in the XRD peak positions of NCs relative to the bulk references are similar to the shifts previously reported for nitrides due to a slight nitrogen excess or deficiency[16,17]. The reaction between Ta(V) halides and NH$_3$ yielded an orthorhombic Ta$_3$N$_5$ phase instead of cubic TaN (**Fig. 2f**). Most of the Ta$_3$N$_5$ NCs exhibit anisotropic growth, appearing as rodlike structures. W$_2$N NCs can also be produced by the same method using WCl$_6$ as the precursor (Supplementary Fig. 10). All of these materials formed stable colloidal solutions, corroborated by SAXS and dynamic light scattering measurements (Supplementary Figs. 5 and 11). In addition to binary nitride phases, our high-pressure molten salt synthetic method can be expanded to ternary nitrides. For example, Ti$_{1-x}$V$_x$N NCs were directly synthesized using mixtures of TiI$_4$ and VCl$_3$. Powder XRD patterns revealed a gradual shift of all diffraction peaks in Ti$_{1-x}$V$_x$N NCs in proportion to the increased V content and TEM images showed crystalline ternary nitride NCs (Supplementary Fig. 8). Both GaN and TMN NCs show excellent stability against oxidation, even after a prolonged (24 hours) heating at 150 °C in air (Supplementary Fig. 12). This observation further supports the utility of metal nitride NCs for real-world applications.

The synthesis of VN NCs can be described as a metathesis reaction: $VCl_3 + 4NH_3 \rightarrow VN + 3NH_4Cl$, but syntheses of other nitride NCs can involve redox reactions where metals change their oxidation state, as confirmed by X-ray photoelectron spectroscopy (XPS, Supplementary Figs. 13 to 18, Supplementary Tables 2 to 7). $NH_3$ can participate in the redox chemistry, either reducing to $N^{3-} + H_2$ or oxidizing to $N_2 + H^+$. For example, in GaN NC synthesis, we found that a metathesis reaction using $GaI_3$ and $NH_3$ did not produce any crystalline GaN (Supplementary Fig. 19). However, the reaction of $NH_3$ with a reduced gallium halide, $Ga_2I_4$ containing $Ga^I$ and $Ga^{III}$ produces highly crystalline GaN (**Fig. 2c** and Supplementary Fig.7). Gas chromatography of the reactor head space revealed the presence of $H_2$ (Supplementary Fig. 19c), indicating that reduced Ga halides are simultaneously Ga sources and reductants that activate $NH_3$ during the formation of GaN NCs. Redox reactions are also indicated by the products of the reactions between $Ti^{IV}$, $Nb^V$, $Mo^V$, $W^{VI}$ halides and $NH_3$ in molten salts, whereas $Ta^V$ showed less tendency to be reduced, probably because of the stability of the $Ta^V$ oxidation state. Similar redox reactions were reported in the CVD growth of nitride films using metal halides and $NH_3$[11,12].

The pressures used in this work are insufficient to change the thermodynamic characteristics of the solid phases. However, pressure can impact the kinetics of NC nucleation and growth, either through the activation volume, a pressure analog of activation energy in the Arrhenius equation[37], or through altering the concentrations of the nitrogen species ($NH_3$, $NH_2^-$, *etc.*) present to the reaction mixture. To the best of our knowledge, no experimental information is available on how pressure impacts the kinetics of nucleation and growth of colloidal NCs. To test the first hypothesis, we carried out a series of control experiments where we ran colloidal syntheses of NCs under argon pressures ranging from 0.1 to 8.6 MPa (Supplementary Note 1). Somewhat to our surprise, the effect of gas pressure was negligible on the size and size distribution of the model NC system studied (PbS NCs). It affirms that the activation volumes of NC nucleation and growth are too small to impact the synthesis by pressure alone.

Metal nitrides synthesized in a molten salt at 100 kPa of $NH_3$ pressure typically appear as aggregates or irregularly shaped particles formed through oriented attachment and sintering of small NCs (**Fig. 3a** and Supplementary Figs. 20 to 22), while an $NH_3$ pressure of 2.0 MPa under otherwise identical conditions yields well-separated cubic NCs terminated with non-polar {100} facets (**Fig. 3b**). This observation suggests a relation between ammonia pressure, aggregation and oriented attachment of nitride NCs. To aggregate, the NCs must overcome repulsive barriers responsible for colloidal stability. Since colloidal stability in molten salts is related to surface-templated layering of the ions around each NC, the presence of surface-binding ions is critical for the stabilization of colloidal dispersions[28]. In iodide molten salts, Lewis-acidic metal sites present at the NC surface can interact with Lewis basic moieties including $I^-$, $NH_3$, and $NH_2^-$.

Our *ab initio* molecular dynamics (AIMD) simulations showed that non-polar {100} surfaces of a VN NC exhibit only a weak affinity toward $I^-$ ions when immersed in molten KI. In contrast, rapid adsorption of $NH_3$ and $NH_2^-$ onto the V sites of the non-polar {100} VN surface was observed at picosecond time scale (Supplementary Fig. 23). Once bound, the nitrogen species are thermodynamically trapped on VN surfaces against further movement. We further computed the binding energies of $NH_3$ and $NH_2$ on the VN {100} surface as a function of the coverage (**Fig. 3c**) by density function theory (DFT), which suggests the very favorable interaction of $NH_2$ on the VN surface up to 100% coverage of metal surface sites (Supplementary Fig. 24); we also found the synergistic effect of $NH_3$-$NH_2$ co-adsorption due to the formation of hydrogen bonding between $NH_3$ and $NH_2$ on the VN surface (Supplementary Fig. 25). Fourier transform infrared (FTIR) spectroscopy revealed the presence of N-H stretching vibrations in as-synthesized VN particles (**Fig. 3d**), confirming the presence of $NH_3$ and $NH_2$ surface terminations. Similar binding trends were also observed on rock salt TiN {100} surface and zinc blende GaN {111} surface (Supplementary Fig. 26). The surface binding of $NH_3$ and $NH_2^-$ creates a nitrogen-rich shell that protects NCs dispersed in a molten salt from direct contact. Furthermore, the charges of the surface-bound $NH_2^-$ ions template the layering of molten salt ions (**Fig. 3e,f**) responsible for colloidal stability and preventing the aggregation of nitride NCs[28].

The solubility of $NH_3$ in molten salts[38], e.g., $KNO_3$/$NaNO_3$/$LiNO_3$ eutectic, is ~0.2 $mol\ L^{-1}\ MPa^{-1}$, and a saturated solution at 100 kPa partial pressure contains only ~0.02 $mol\ L^{-1}$ of $NH_3$, which is much smaller than the ~1 $mol\ L^{-1}$ initial concentration of reactive metal ions dissolved in the molten salt. Dissolved $NH_3$ partially dissociates forming $NH_2^-$ and $NH_4^+$ ions, with an overall dissociation energy of 81.2 kJ/mol for $2NH_3 \rightarrow NH_2^- + NH_4^+$ reaction[39]. The equilibrium concentrations of $NH_4^+$ and $NH_2^-$ ions are lower than ~$10^{-5}\ mol\ L^{-1}$ at 500 °C and 100 kPa $NH_3$ pressure (Supplementary Note 2). The above estimate shows that, at an ambient ammonia pressure, nitride NCs nucleate and grow under a strong deficiency of nitrogen precursors.

Such conditions cannot provide a sufficient coverage of surface-bound $NH_3$ and $NH_2^-$ ions to establish colloidal stability and prevent aggregation of growing nitride NCs (Supplementary Fig. 27). It has been also demonstrated that an excess of ammonia is helpful for growing defect-free nitride crystals, *e.g.*, CVD synthesis of GaN uses > 20:1 ratio of $NH_3$-to-$GaCl_3$ at 700-1100 °C[10,40]. When the ammonia pressure is 2.0 MPa, the equilibrium concentration of dissolved $NH_3$ increases proportionally with pressure to ~0.4 $mol\ L^{-1}$, which is comparable to the concentration of metal precursors, and falls into the range of conditions used in traditional colloidal syntheses[16,41]. When the ammonia pressure is above the blue region in **Fig. 1a**, the crystal growth proceeds too rapidly toward larger crystals.

The metal nitrides reported in this study represent technologically important semiconductors (GaN), superconductors (NbN), and plasmonic materials (TiN, VN). By exploring synthetic conditions previously inaccessible for colloidal NCs, we aim to access materials with novel properties. For example, extensive studies of GaN synthesized by CVD and MBE methods have revealed a strong relation between growth temperature and luminescence – a growth or annealing temperature above 700 °C was required for observing a strong band-edge photoluminescence (PL) in GaN films, while the materials synthesized at lower temperatures showed trap emission because of mid-gap states introduced by gallium and nitrogen vacancies[42,43]. A similar behavior was observed for colloidal GaN NCs synthesized in this study – all samples synthesized below 500 °C showed broad yellow-red emission from trap states, however, starting from about 525 °C, GaN NCs showed predominantly band-edge ultraviolet PL (**Fig. 4a,b** and Supplementary Figs. 28 to 30). Further spectroscopic studies (transient absorption, PL lifetime) support the assignment of PL band to band-edge GaN emission (Supplementary Note 3, Supplementary Figs. 31 to 33). The transition from trap- to band-edge emission shows no obvious correlation with TEM and XRD data (Supplementary Fig. 29), but the Raman spectra clearly correlate with the emergence of band-edge PL (**Fig. 4c**). Samples synthesized at 525 °C and above have well-defined $A_1$(LO) and $E_2$(high) GaN phonon modes and corresponding overtone peaks closely resembling the Raman spectrum of CVD-grown GaN (Supplementary Fig. 30). In contrast, GaN NCs synthesized below 500 °C show very weak and broadened Raman peaks, characteristic of disorder associated with vacancy defects in analogy with GaN grown by MBE or CVD at low temperatures[44,45].

TMNs are known for plasmonic properties such as localized surface plasmon resonance (LSPR), which resemble plasmonic properties of noble metal NCs[46]. The extinction spectrum of a colloidal solution of TiN NCs shows the LSPR peak at 757 nm (**Fig. 4d**), which overlaps with the biological transparency window previously identified as the requirement for theragnostic applications of plasmonic nanomaterials[47]. The synthetic availability of colloidal TiN, combined with the biocompatibility of TiN which is widely used in medical implants[48], can offer new opportunities in biomedical applications such as photothermal therapy. Alloying vanadium with a larger number of d-electrons shifts LSPR of $Ti_{1-x}V_xN$ NCs to higher frequencies toward LSPR at 550 nm for VN NCs (**Fig. 4d**). Many TMNs, *e.g.*, NbN and VN, also exhibit superconductivity with higher critical temperatures and superior chemical stability compared to bare metal superconductors and can be functionally integrated with nitride semiconductor devices[49]. Magnetic susceptibility measurements of NbN NCs revealed the onset of diamagnetism below 16 K, which we attributed to the Meissner effect (Supplementary Fig. 34)[50].

In conclusion, we want to emphasize that this work just scratches the surface of colloidal nitrides that can be synthesized in molten inorganic salts. Our results have demonstrated a general route toward colloidal metal nitride NCs using molten halide salts and ammonia, where $NH_3$ pressure allows controlling the reaction product morphology by stabilizing colloidal nitride NCs against aggregation and sintering. Naturally, we anticipate the emergence of colloidal nitride QDs for optoelectronic applications, backed by the prominence of GaN semiconductor family. Beyond the nitrides, further development of synthetic methodology towards colloidal NCs of other refractory materials with covalent bonds would be highly desired.

**Acknowledgements**

The authors thank Prof. Henry La Pierre and Grant Wilkinson for helping with SQUID measurements, Dr. Shengsong Yang, Benjamin Hammel, and Prof. Gordana Dukovic for helping with STEM-EDS measurements, Dr. Yingjie Fan for helping with gas chromatography measurements, Dr. Minhal Hasham for helping with PL lifetime measurements, and Dr. James Weng for helping with X-ray PDF measurements at the early stage of the project. We thank Dr. Jun Hyuk Chang, Dr. Aritrajit Gupta, and Yi-Chen Chen for stimulating discussions. We are particularly grateful to Dr. Andrew Nelson for a critical reading and editing of the manuscript.

**Funding:** The work on synthesis of colloidal GaN NCs was funded by the U.S. Department of Energy, Office of Science, Basic Energy Sciences, Materials Sciences and Engineering Division, under grant DE-SC0025256 and by the Samsung QD Cluster Collaboration. The work on synthesis of transition metal nitride NCs and DFT computation was supported by the U.S. National Science Foundation under Grant Number CHE-2318105 (M-STAR CCI). Spectroscopic studies of nitride NCs were supported by the Department of Defense Air Force Office of Scientific Research under grant number FA9550-22-1-0283. The mechanistic studies of high-pressure colloidal synthesis were supported by the National Science Foundation Science and Technology Center (STC) for Integration of Modern Optoelectronic Materials on Demand (IMOD) under Award No. DMR-2019444. This work made use of the shared facilities at the University of Chicago Materials Research Science and Engineering Center, supported by National Science Foundation under award number DMR-2011854. Parts of this work were carried out at the Soft Matter Characterization Facility of the University of Chicago. Work performed at the Center for Nanoscale Materials, a U.S. Department of Energy Office of Science User Facility, was supported by the U.S. DOE, Office of Basic Energy Sciences, under Contract No. DE-AC02-06CH11357.

**Author Contributions:** R.L. conceived, designed, and performed experiments, analyzed data, and co-wrote the manuscript. V.K. performed AIMD and DFT simulations under supervision by D.J. N.J. contributed to magnetic property tests on nitride NCs and data analysis under supervision by J.S.A. W.C. contributed to the development of high-pressure synthesis and building of the high-pressure setup. Z.Z. supported nitride synthesis using ambient-pressure $NH_3$ and built the ambient-pressure molten salt synthesis setup. D.W. performed X-ray characterization and XPS data analysis. J.C.O. contributed to the development of molten salt redox chemistry and carried out transient absorption data analysis. Z.M. carried out the purification of nitride colloids and SAXS data analysis. J.C. contributed to the preparation of salt precursors. A.M.H. investigated colloidal PbS NC synthesis under elevated argon pressure. A.S.F. performed XPS measurements. R.D.S. performed femtosecond transient absorption spectroscopy studies. D.V.T. conceived and supervised the project, acquired funding, analyzed data, and co-wrote the manuscript. All authors discussed the results and commented on the manuscript.

**Competing interests:** R.L. and D.V.T. have filed provisional patents related to this work.

# References

1. Nakamura, S., Mukai, T. & Senoh, M. Candela-class high-brightness InGaN/AlGaN double-heterostructure blue-light-emitting diodes. *Appl. Phys. Lett.* **64**, 1687-1689 (1994).

2. Amano, H. et al. The 2018 GaN power electronics roadmap. *J. Phys. D: Appl. Phys.* **51**, 163001 (2018).

3. Höhn, P. & Niewa, R. Nitrides of non-main group elements. in *Handbook of Solid State Chemistry, Part 1* (Wiley-VCH, 2017).

4. Sun, W. et al. A map of the inorganic ternary metal nitrides. *Nat. Mater.* **18**, 732-739 (2019).

5. Gao, Z. et al. Shielding Pt/γ-$Mo_2N$ by inert nano-overlays enables stable $H_2$ production. *Nature* **638**, 690–696 (2025).

6. Hashimoto, T. et al. A GaN bulk crystal with improved structural quality grown by the ammonothermal method. *Nat. Mater.* **6**, 568-571 (2007).

7. Wang, D. et al. Ferroelectric YAlN grown by molecular beam epitaxy. *Appl. Phys. Lett.* **123**, 033003 (2023).

8. Skidmore, C. H. et al. Proximity ferroelectricity in wurtzite heterostructures. *Nature* **637**, 574–579 (2025).

9. Talley, K. R. et al. Synthesis of $LaWN_3$ nitride perovskite with polar symmetry. *Science* **374**, 1488-1491 (2021).

10. Kuykendall, T. et al. Complete composition tunability of InGaN nanowires using a combinatorial approach. *Nat. Mater.* **6**, 951-956 (2007).

11. Fix, R., Gordon, R. G. & Hoffman, D. M. Chemical vapor deposition of titanium, zirconium, and hafnium nitride thin films. *Chem. Mater.* **3**, 1138-1148 (1991).

12. Fix, R., Gordon, R. G. & Hoffman, D. M. Chemical vapor deposition of vanadium, niobium, and tantalum nitride thin films. *Chem. Mater.* **5**, 614-619 (1993).

13. Talapin, D. V. et al. Prospects of colloidal nanocrystals for electronic and optoelectronic applications. *Chem. Rev.* **110**, 389–458 (2010).

14. García de Arquer, F. P. et al. Semiconductor quantum dots: Technological progress and future challenges. *Science* **373**, eaaz8541 (2021).

15. Wang, H. et al. Transition metal nitrides for electrochemical energy applications. *Chem. Soc. Rev.* **50**, 1354-1390 (2021).

16. Xu, X. et al. Two-dimensional arrays of transition metal nitride nanocrystals. *Adv. Mater.* **31**, 1902393 (2019).

17. Guy, K. et al. Original synthesis of molybdenum nitrides using metal cluster compounds as precursors: Applications in heterogeneous catalysis. *Chem. Mater.* **32**, 6026-6034 (2020).

18. Karaballi, R. A. et al. Synthesis of plasmonic group-4 nitride nanocrystals by solid-state metathesis. *Angew. Chem. Int. Ed.* **58**, 3147-3150 (2019).

19. Giordano, C. et al. Metal nitride and metal carbide nanoparticles by a soft urea pathway. *Chem. Mater.* **21**, 5136-5144 (2009).

20. Murray, C. B., Norris, D. J. & Bawendi, M. G. Synthesis and characterization of nearly monodisperse CdE (E= sulfur, selenium, tellurium) semiconductor nanocrystallites. *J. Am. Chem. Soc.* **115**, 8706-8715 (1993).

21. Yin, Y. & Alivisatos, A. P. Colloidal nanocrystal synthesis and the organic–inorganic interface. *Nature* **437**, 664-670 (2005).

22. Parvizian, M. & De Roo, J. Precursor chemistry of metal nitride nanocrystals. *Nanoscale* **13**, 18865-18882 (2021).

23. Yang, L. et al. Cation exchange in colloidal transition metal nitride nanocrystals. *J. Am. Chem. Soc.* **146**, 12556–12564 (2024).

24. Vaughn II, D. D. et al. Solution synthesis of $Cu_3PdN$ nanocrystals as ternary metal nitride electrocatalysts for the oxygen reduction reaction. *Chem. Mater.* **26**, 6226–6232 (2014).

25. Shanker, G. S. & Ogale, S. Faceted colloidal metallic $Ni_3N$ nanocrystals: size-controlled solution-phase synthesis and electrochemical overall water splitting. *ACS Appl. Energy Mater.* **4**, 2165-2173 (2021).

26. Taylor, P. N. et al. Synthesis of widely tunable and highly luminescent zinc nitride nanocrystals. *J. Mater. Chem. C* **2**, 4379 (2014).

27. Lange, N. A. *Lange's Handbook of Chemistry* (McGraw-Hill, New York, NY, 15th ed., 1999).

28. Zhang, H. et al. Stable colloids in molten inorganic salts. *Nature* **542**, 328-331 (2017).

29. Srivastava, V. et al. Colloidal chemistry in molten salts: Synthesis of luminescent $In_{1-x}Ga_xP$ and $In_{1-x}Ga_xAs$ quantum dots. *J. Am. Chem. Soc.* **140**, 12144-12151 (2018).

30. Zhou, Z. et al. Colloidal chemistry in molten inorganic salts: Direct synthesis of III–V quantum dots via dehalosilylation of $(Me_3Si)_3Pn$ (Pn = P, As) with group III halides. *J. Am. Chem. Soc.* **147**, 9198–9209 (2025).

31. Ondry, J. et al. Reductive pathways in molten inorganic salts enable colloidal synthesis of III-V semiconductor nanocrystals. *Science* **386**, 401-407 (2024).

32. Portehault, D. et al. A general solution route toward metal boride nanocrystals. *Angew. Chem. Int. Ed.* **50**, 3179-3182 (2011).

33. Liu, X., Fechler, N. & Antonietti, M. Salt melt synthesis of ceramics, semiconductors and carbon nanostructures. *Chem. Soc. Rev.* **42**, 8237-8265 (2013).

34. Guan, H. et al. General molten-salt route to three-dimensional porous transition metal nitrides as sensitive and stable Raman substrates. *Nat. Commun.* **12**, 1376 (2021).

35. Cho, W. et al. Synthesis of colloidal GaN and AlN nanocrystals in biphasic molten salt/organic solvent mixtures under high-pressure ammonia. *ACS Nano* **17**, 1315-1326 (2023).

36. Parvizian, M. et al. Molten salt-assisted synthesis of titanium nitride. *Small Methods* **8**, 2400228 (2024).

37. Jacobs, K. et al. Activation volumes for solid-solid transformations in nanocrystals. *Science* **293**, 1803–1806 (2001).

38. Allulli, S. Solubilities of ammonia in alkali nitrate and perchlorate melts. *J. Phys. Chem.* **73**, 1084–1087 (1969).

39. Jolly, W. A. Heats, free energies, and entropies in liquid ammonia. *Chem. Rev.* **50**, 351–361 (1952).

40. Takekawa, N. et al. GaN growth via tri-halide vapor phase epitaxy using solid source of $GaCl_3$: Investigation of the growth dependence on $NH_3$ and additional $Cl_2$. *Jpn. J. Appl. Phys.* **58**, SC1022 (2019).

41. Liu, Q. et al. Theory-guided synthesis of highly luminescent colloidal cesium tin halide perovskite nanocrystals. *J. Am. Chem. Soc.* **143**, 5470–5480 (2021).

42. Nakamura, S. et al. Thermal annealing effects on p-type Mg-doped GaN films. *Jpn. J. Appl. Phys.* **31**, L139 (1992).

43. Jain, S. C. et al. III–nitrides: Growth, characterization, and properties. *J. Appl. Phys.* **87**, 965–1006 (2000).

44. Yu, K. M. et al. Effects of native defects on properties of low-temperature grown, non-stoichiometric gallium nitride. *J. Phys. D: Appl. Phys.* **48**, 385101 (2015).

45. Hubáček, T. et al. Improvement of luminescence properties of GaN buffer layer for fast nitride scintillator structures. *J. Cryst. Growth* **464**, 221–225 (2017).

46. Guler, U., Shalaev, V. M. & Boltasseva, A. Nanoparticle plasmonics: Going practical with transition metal nitrides. *Mater. Today* **18**, 227–237 (2015).

47. Tsai, M.-F. et al. Au nanorod design as light-absorber in the first and second biological near-infrared windows for in vivo photothermal therapy. *ACS Nano* **7**, 5330–5342 (2013).

48. van Hove, R. P. et al. Titanium-nitride coating of orthopaedic implants: A review of the literature. *BioMed Res. Int.* **2015**, 485975 (2015).

49. Yan, R. et al. GaN/NbN epitaxial semiconductor/superconductor heterostructures. *Nature* **555**, 183–189 (2018).

50. Zolotavin, P. & Guyot-Sionnest, P. Meissner effect in colloidal Pb nanoparticles. *ACS Nano* **4**, 5599–5608 (2010).

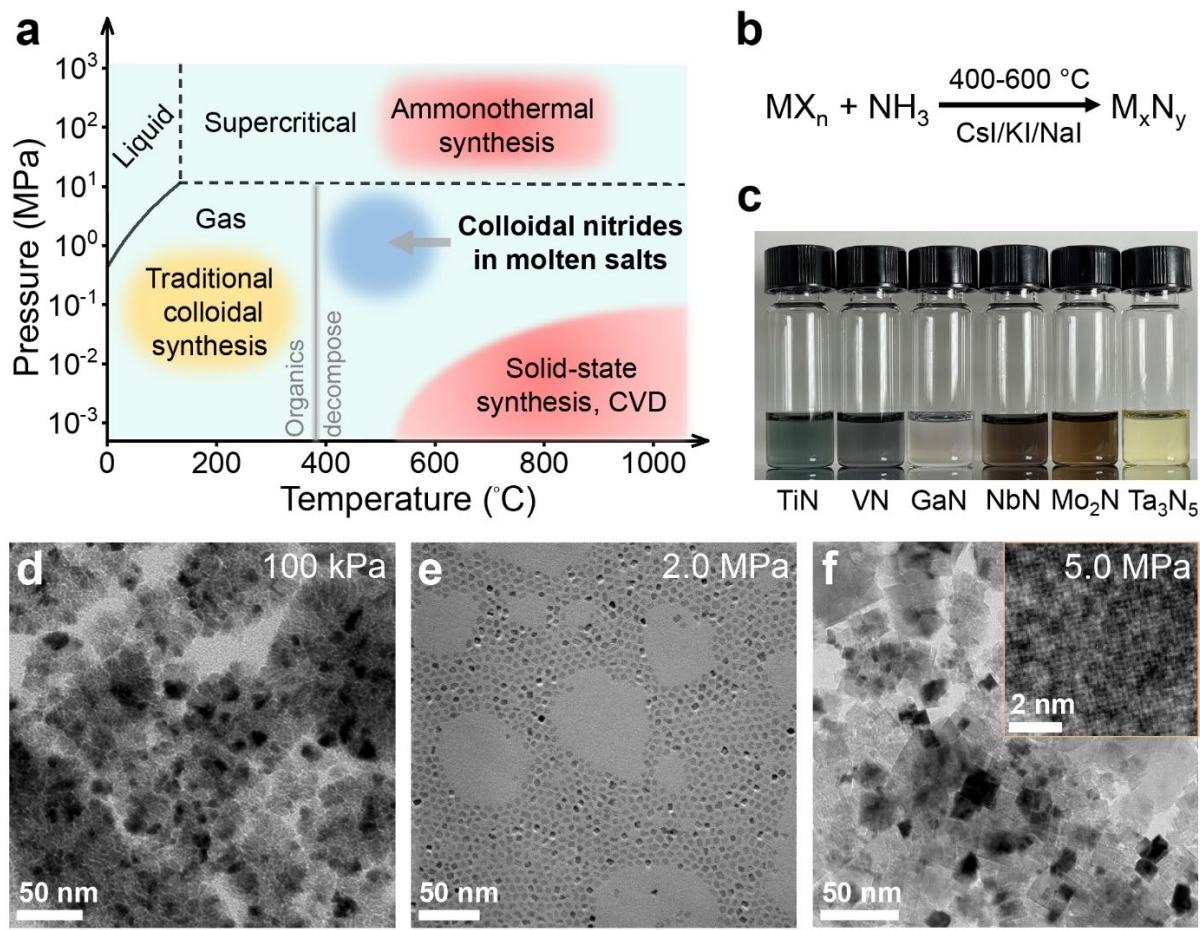

**Fig. 1. Effect of NH₃ pressure on morphology of metal nitrides synthesized in molten inorganic salts. (a)** Parameter space accessible for molten salt colloidal nitrides (marked as blue region) compared with traditional colloidal synthesis of nanocrystals (NCs), ammonothermal synthesis, CVD, and solid-state syntheses of nitride crystals and films. **(b)** Reactions used for synthesis of nitride NCs. **(c)** Photograph of colloidal solutions of different metal nitride NCs synthesized under conditions marked as the blue region in panel (a). All NCs are capped with oleate/oleylamine ligands and dispersed in toluene. **(d-f)** TEM images of VN products synthesized using 0.1, 2.0, and 5.0 MPa NH₃ pressure, respectively.

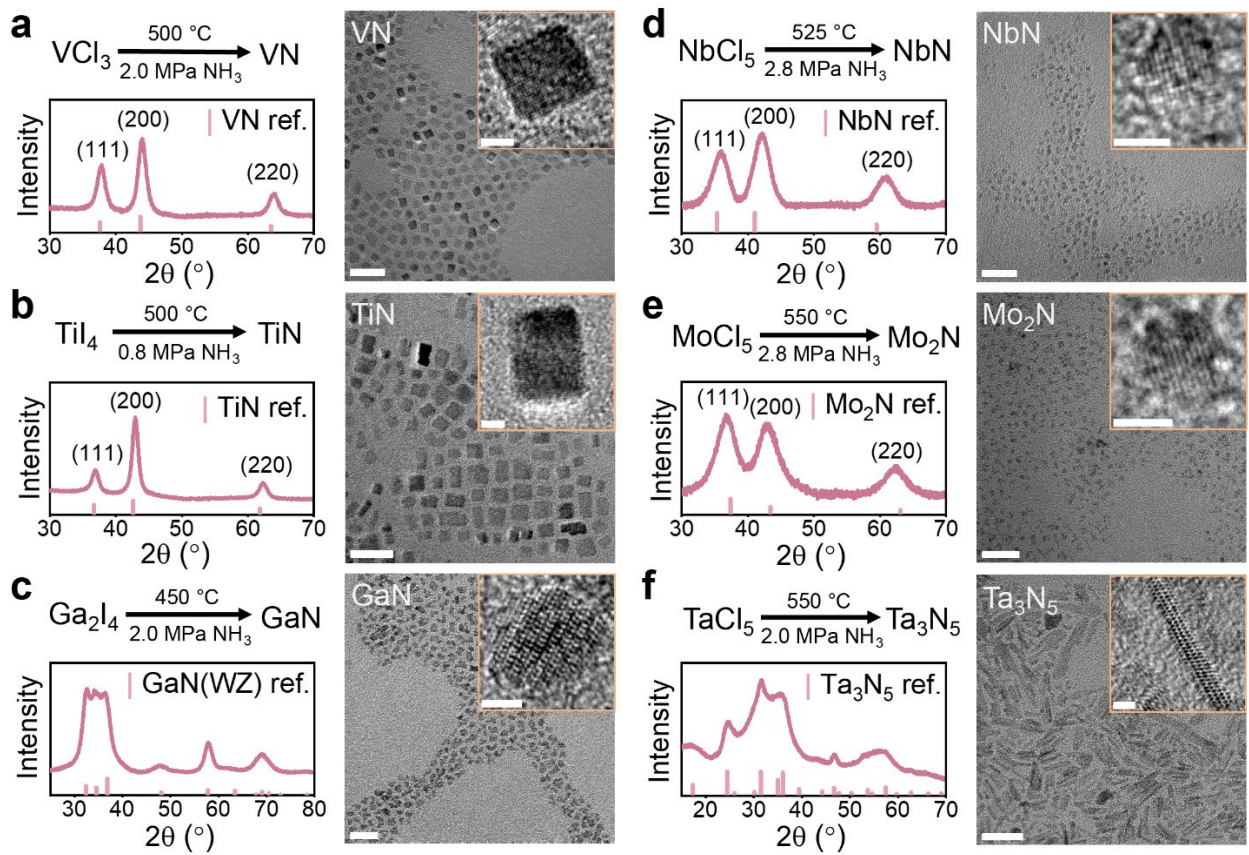

**Fig. 2. Colloidal metal nitride nanocrystals synthesized in molten salts.** Reaction schemes, XRD patterns, and TEM images (low resolution and, inset, high resolution) of **(a)** VN NCs. **(b)** TiN NCs. **(c)** GaN NCs. **(d)** NbN NCs. **(e)** $Mo_2N$ NCs. **(f)** $Ta_3N_5$ NCs. Scale bars for low-resolution and high-resolution TEM images and are 20 nm and 2 nm, respectively.

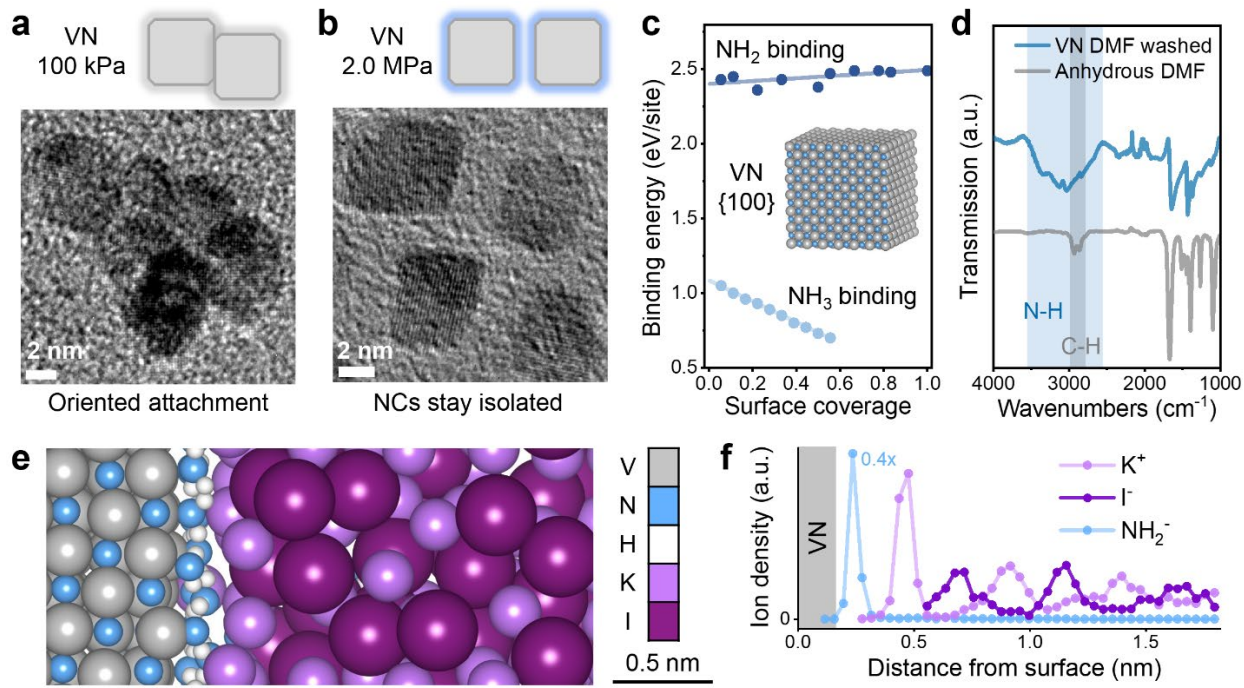

**Fig. 3. Colloidal stabilization of nitride nanocrystals in molten salts. (a,b)** Schematic demonstration (top) and high-resolution TEM images (bottom) of VN NCs synthesized at 500°C using (a) 100 kPa $NH_3$ and (b) 2.0 MPa $NH_3$ pressure. **(c)** DFT-calculated binding energies of $NH_3$ and $NH_2$ on VN {100} facets with different surface coverage. (inset) Sketch of a cubic VN NC terminated with {100} non-polar facets. **(d)** FTIR spectra of as-synthesized VN NCs rinsed with anhydrous N,N-dimethylformamide (DMF). The broad absorption band at 2600-3500 cm$^{-1}$ corresponds to N-H stretching vibrations of molecules and ions bound to VN NC surface and possible C-H stretching vibrations from trace amount of residual DMF. **(e)** A snapshot from *ab initio* Molecular Dynamics simulations of the interface between the $NH_2^-$ covered VN {100} surface of a cubic VN NC and molten KI. **(f)** Number density profiles of $K^+$ and $I^-$ ions near the interface, showing strong ordering of the molten salt templated by the adsorbed $NH_2^-$ layer on the NC surface. The distance is measured from the averaged position of all the V and N atom centers in the top layer of the VN {100} surface.

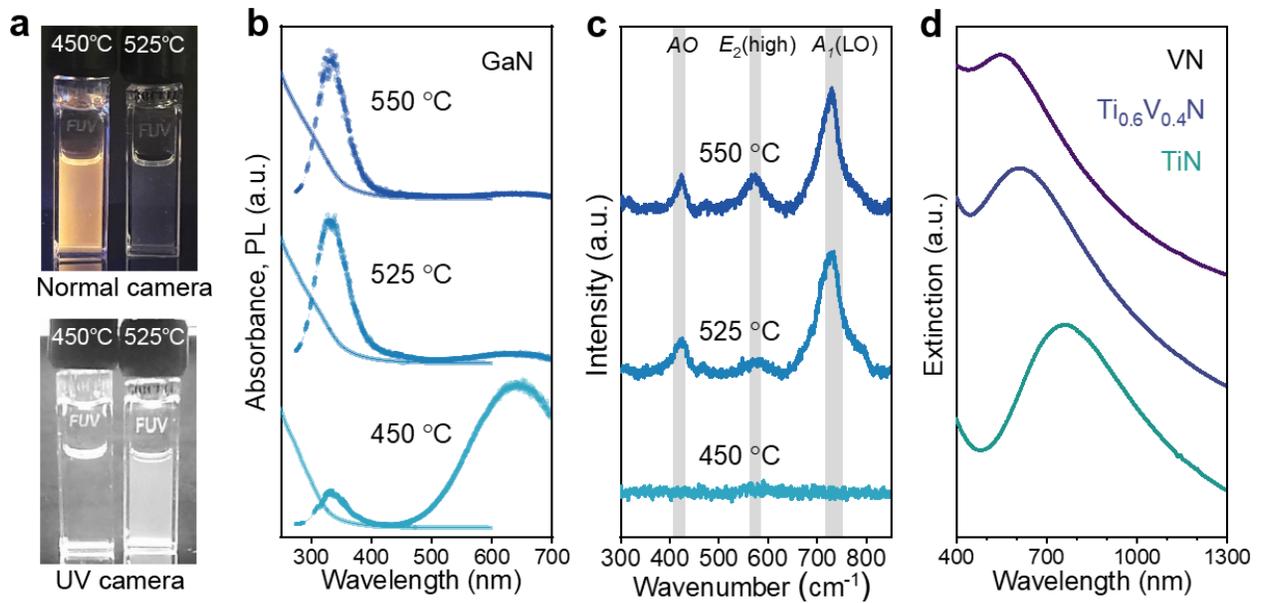

**Fig. 4. Emissive and plasmonic properties of colloidal nitride nanocrystals. (a)** Photographs of colloidal GaN NCs synthesized at 450°C and 525°C, dispersed in methylcyclohexane and illuminated with UV light (254 nm). The photographs capture (top) visible 400 nm – 700 nm emission and (bottom) UV 290 nm – 400 nm emission. **(b)** Absorption (solid line) and room-temperature PL spectra (dots) of colloidal GaN NCs synthesized at 450°C, 525°C and 550°C. The left branch of PL was fit to a Gaussian function (dash line). **(c)** Raman spectra of colloidal GaN NCs with reference GaN Raman modes indicated as grey lines. **(d)** Extinction spectra of colloidal TiN, $Ti_{0.6}V_{0.4}N$, and VN NCs in toluene showing localized surface plasmon resonance at 757 nm, 612 nm, and 550 nm, respectively.